# The Truth About Ballistic Coefficients

Michael Courtney, PhD, Ballistics Testing Group, PO Box 3101, Cullowhee, NC, 28723
Michael_Courtney@alum.mit.edu

Amy Courtney, PhD, Department of Engineering, Western Carolina University, Cullowhee, NC, 28723  Amy_Courtney@post.harvard.edu

**Abstract:**
The ballistic coefficient of a bullet describes how it slows in flight due to air resistance. This article presents experimental determinations of ballistic coefficients showing that the majority of bullets tested have their previously published ballistic coefficients exaggerated from 5-25% by the bullet manufacturers.  These exaggerated ballistic coefficients lead to inaccurate predictions of long range bullet drop, retained energy and wind drift.

**Introduction**
The ballistic coefficient (BC) describes how air resistance slows a projectile in flight [SPE94, BAR97, HOR91, NOS96].  Accurate quantification of BC can be important in predicting long-range bullet drop, wind drift, and retained energy.  The models and equations describing how BC determines velocity loss over flight distance are well known [RES06].  To a rough approximation, the BC can be estimated as the fraction of 1000 yards over which a projectile loses half of its initial kinetic energy.  In other words, a bullet with a BC of 0.300 should lose roughly half of its initial kinetic energy at a range of 300 yards.

However, many bullet manufacturers exaggerate BC specifications for marketing purposes because BC is perceived to be important by customers, and because many manufactures rely on overly optimistic theoretical predictions that ignore the effects of the engraving of rifling, manufacturing defects, imperfect alignment of bullet axis and velocity vector, and other factors.

This paper presents the results of careful BC measurements for fourteen bullets.  The test bullets represent five manufacturers, a weight range of 40-220 grains, calibers from 0.224 to .308, and published BC values between 0.200 and 0.523.  Results (*Table 1*) show that manufacturer claims regarding bullet BC are often exaggerated.  This exaggeration leads to inaccurate predictions of bullet drop, wind drift, impact energy, and impact velocity.

**Method**
Two chronographs are used to measure near and far bullet velocities at distances of 8 feet and 299 feet from the muzzle.  The velocity loss over the separation distance of 291 feet is used along with the relative humidity, air temperature, atmospheric pressure, and altitude to compute the bullet BC using the G1 resistance model [JBM07].  The BC is determined individually for three to six separate shots, and these BC measurements are used to compute the mean measured BC and estimate the measurement uncertainty due to shot-to-shot variations.

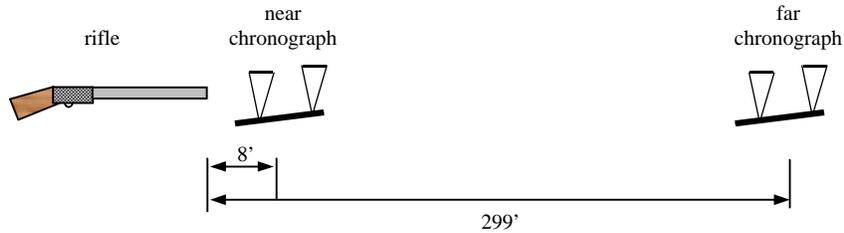

**Results**
The table compares the results of careful BC measurements with the manufacturers' claims. Numbers in parentheses represent the estimated uncertainty in the last significant digit(s) of the measured BC.

*Table 1:*

| Manufacturer | Caliber (inches) | Weight (grains) | Style | Published BC | Reference | Measured BC | Exaggeration (percent) | Near Velocity (fps) |
|---|---|---|---|---|---|---|---|---|
| Hornady | 0.224 | 40 | VMAX | 0.200 | [HOR07] | 0.199(1) | 0.50% | 3137 |
| Hornady | 0.224 | 55 | SP | 0.235 | [HOR91] | 0.218(3) | 7.80% | 2400 |
| Barnes | 0.224 | 53 | XFB | 0.231 | [BAR97] | 0.197(10) | 17.26% | 2874 |
| Nosler | 0.308 | 150 | BT | 0.435 | [NOS96] | 0.381(7) | 14.17% | 2570 |
| Hornady | 0.308 | 150 | FMJBT | 0.398 | [HOR91] | 0.361(23) | 10.25% | 2656 |
| Winchester | 0.308 | 168 | CTBST | 0.475 | [WIN07] | 0.421(4) | 12.83% | 2644 |
| Hornady | 0.308 | 110 | VMAX | 0.290 | [HOR07] | 0.247(28) | 17.41% | 3501 |
| Nosler | 0.308 | 125 | BT | 0.366 | [NOS96] | 0.306(5) | 19.61% | 2245 |
| Nosler | 0.308 | 125 | BT | 0.366 | [NOS96] | 0.308(10) | 18.83% | 2794 |
| Nosler | 0.308 | 125 | BT | 0.366 | [NOS96] | 0.319(11) | 14.73% | 3010 |
| Barnes* | 0.308 | 150 | TSX | 0.428 | [BAR05] | 0.349(20) | 22.64% | 2567 |
| Hornady | 0.308 | 150 | RN | 0.186 | [HOR91] | 0.163(6) | 14.11% | 2624 |
| Hornady | 0.308 | 165 | SPBT | 0.435 | [HOR91] | 0.406(30) | 7.14% | 2750 |
| Hornady | 0.308 | 220 | RN | 0.300 | [HOR91] | 0.249(9) | 20.48% | 2444 |
| Winchester | 0.257 | 85 | CTBST | 0.329 | [NOS07] | 0.309(9) | 6.47% | 3449 |
| Berger | 0.257 | 115 | VLD | 0.523 | [BER07] | 0.419(4) | 24.82% | 3148 |

**Discussion and Conclusion**
An example demonstrates the implications of an exaggerated ballistic coefficient for the Berger 115 grain VLD shot at 3148 fps. For a zero range of 200 yards and a 10 mph cross wind, using the manufacturer's claimed ballistic coefficient of 0.523 gives a drop of 45.1 in, a wind drift of 19.6 in, and an impact energy of 1180 ft-lbs at 550 yards for atmospheric conditions of 30º F, 0% relative humidity, and 29.92 mm Hg.  In contrast, using the more accurate BC of 0.419 gives a drop of 49.8 in, a wind drift of 25.7 in, and impact energy of 956 ft-lbs under the same conditions.

---
* Barnes has recently undertaken to more carefully determine ballistic coefficients and have measured the BC of this bullet to be 0.369 [BAR07].

Errors in trajectory and wind drift predictions lead to the point of impact being different from expectations.  Errors in impact velocity predictions can cause unexpected failures in bullet performance, because many bullet designs have a window of impact velocities over which they expand reliably.  Projectiles impacting below a threshold velocity can result in failure to expand and sub-optimal terminal performance.  Errors in impact energy predictions lead to overly optimistic expectations regarding terminal performance.

In conclusion, manufacturers' published values for BC are exaggerated for many bullets, some by nearly 25%.  Bullets also exhibit shot-to-shot variations of 1-5% in BC that suggest an inherent accuracy limit in predictions based on BC measurements.  It should be noted that the BC can depend on the muzzle velocity and on the particular firearm used.  If a BC is needed with less than 5% uncertainty, the BC should be determined with the same firearm.